\documentclass[prb,reprint]{revtex4-1}
\pdfoutput=1
\usepackage{graphicx}
\usepackage{color}
\usepackage{bm}% bold math
\usepackage{hyperref}% add hypertext capabilities
\begin{document}

\title{Unusual Landau Level Pinning and Correlated $\nu=1$ Quantum Hall Effect in Hole Systems Confined to Wide GaAs Quantum Wells}
\date{today}

\author{Yang Liu}
\affiliation{Department of Electrical Engineering,
Princeton University, Princeton, New Jersey 08544}
\author{S.\ Hasdemir}
\affiliation{Department of Electrical Engineering,
Princeton University, Princeton, New Jersey 08544}
\author{M.\ Shayegan}
\affiliation{Department of Electrical Engineering,
Princeton University, Princeton, New Jersey 08544}
\author{L.N.\ Pfeiffer}
\affiliation{Department of Electrical Engineering,
Princeton University, Princeton, New Jersey 08544}
\author{K.W.\ West}
\affiliation{Department of Electrical Engineering,
Princeton University, Princeton, New Jersey 08544}
\author{K.W.\ Baldwin}
\affiliation{Department of Electrical Engineering,
Princeton University, Princeton, New Jersey 08544}

\date{\today}

\begin{abstract} 

  In two-dimensional hole systems confined to wide GaAs quantum wells,
  where the heavy- and light- hole states are close in energy, we
  observe a very unusual crossing of the lowest two Landau levels as
  the sample is tilted in magnetic field. At a magic tilt angle
  $\theta\simeq 34^{\circ}$, which surprisingly is independent of the
  well-width or hole density, in a large filling factor range near
  $\nu=1$ the lowest two levels are nearly degenerate as evinced by
  the presence of two-component quantum Hall states. Remarkably, a
  quantum Hall state is seen at $\nu=1$, consistent with a
  \textit{correlated} $\Psi_{111}$ state.

\end{abstract}

\maketitle

\section{Introduction}

Among the most fascinating phases of two-dimensional electron systems
(2DESs) in a strong perpendicular magnetic field ($B_{\perp}$) are the
quantum Hall states (QHSs). These are incompressible phases signaled
by vanishing longitudinal resistance ($R_{xx}$) and quantized Hall
resistance ($R_{xy}$), and are observed at integral or certain
fractional Landau level (LL) filling factors ($\nu$)
\cite{Klitzing.PRL.1980, Tsui.PRL.1982, Jain.CF.2007}. Adding a layer
(or subband) degree of freedom leads to exciting twists. A bilayer
electron system with nearly degenerate LLs from different subbands and
comparable inter- and intra-layer interaction can support new,
two-component (2C) QHSs that have no counterpart in standard
single-layer (or one-component, 1C) 2DESs. An example is the
$\Psi_{331}$ state, a QHS formed at the \textit{even-denominator}
filling $\nu=1/2$ \cite{Suen.PRL.1992, Eisenstein.PRL.1992,
  Suen.PRL.1994, Shabani.PRB.2013}. The correlated $\Psi_{111}$ QHS,
stabilized at $\nu=1$, is another example \cite{Eisenstein.PRL.1992,
  Murphy.PRL.1994, Spielman.PRL.2000, Kellogg.PRL.2004,
  Tutuc.PRL.2004, Eisenstein.ARCMP.2014}. This state is generally
considered to be an excitonic superfluid which can support
Josephson-like interlayer tunneling and superfluid transport.

Recent experimental studies of 2D \textit{hole} systems (2DHSs)
confined to wide GaAs quantum wells (QWs) have unraveled unique
phenomena, arising from the non-trivial spin-orbit coupling of the
heavy- and light-holes. Graninger \textit{et al.} reported a reentrant
behavior of the $\nu=1$ QHS as a function of parallel magnetic field
$B_{||}$ in symmetric, wide QWs \cite{Graninger.PRL.2011}. Later, Liu
\textit{et al.} observed an unusual crossing of the two lowest-energy
LLs at $B_{||}=0$ as a function of $B_{\perp}$ \cite{Liu.PRB.2014}.
For a given density ($p$) and well-width ($W$), the crossing occurs at
a particular filling (Fig. 1(a)); it destroys or weakens the
odd-denominator QHSs near this filling, and stabilizes a unique
even-denominator QHS when it happens at $\nu=1/2$ \cite{Liu.PRB.2014}.

Here we present low-temperature transport data for 2DHSs confined to
symmetric, wide GaAs QWs, as we change the tilt angle ($\theta$)
between the sample normal and the magnetic field direction. We find
that at low and high $\theta$, if $W$ and $p$ are sufficiently large,
LLs from different subbands are well separated from each other and the
2DHSs exhibit normal QHSs at the \emph{standard} fillings $\nu=2/3$,
1, 4/3, 7/5, 8/5 and 5/3. But near an intermediate $\theta$, the 2DHSs
exhibit 2C QHSs similar to those reported in bilayer 2DESs with
vanishing subband separation \cite{Manoharan.PRL.1997}. This
observation indicates that the two lowest-energy LLs are nearly
degenerate and is consistent with a $B_{||}$-induced LL crossing
\cite{Note1}.
Remarkably, as schematically shown in Fig. 1(b), this near degeneracy
persists in a \textit{large magnetic field range} near $\nu=1$ when
$\theta \simeq 34^{\circ}$, a magic angle which does not depend on $W$
or $p$. Moreover, when the two LLs are degenerate, the 2DHS is
compressible at $\nu=1$ if $p$ and $W$ are large so that
$d/l_B\gtrsim 1.3$, but exhibits a QHS when $d/l_B\lesssim 1.3$,
consistent with the development of a correlated 2C ($\Psi_{111}$)
state ($d$ is the interlayer separation and $l_B$ is the magnetic
length) \cite{Murphy.PRL.1994, Tutuc.PRB.2005, Eisenstein.ARCMP.2014}.

\begin{figure}
\includegraphics[width=.45\textwidth]{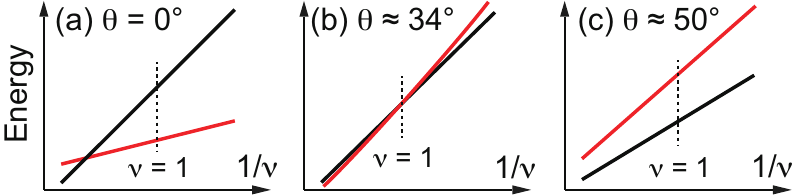}%
\caption{Schematic diagram of the lowest two LLs at different tilt
  angles ($\theta$).}
\end{figure}

\section{Method}

Our samples, grown by molecular beam epitaxy on GaAs (001) wafers,
consist of GaAs QWs flanked by undoped Al$_{0.3}$Ga$_{0.7}$As spacer
and carbon $\delta$-doped layers. The 2DHSs have as-grown densities
ranging from 0.98 to 2.12, in units of $10^{11}$ cm$^{-2}$ which we
use throughout this report, and very high low-temperature mobilities
$\mu \geq$ 100 m$^2$/Vs. We made samples in a van der Pauw geometry, 4
$\times$ 4 mm$^2$, and alloyed In:Zn contacts at their four corners.
Each sample is fitted with an evaporated Ti/Au front-gate and an In
back-gate to control the 2DHS density and QW symmetry. The data
presented here were taken in symmetric QWs. The transport measurements
were carried out in a dilution refrigerator with a base temperature of
$T \approx$ 30 mK and a superconducting magnet up to 18 T. We changed
$\theta$ with an in-situ rotator, and used low-frequency ($\sim 30$
Hz) lock-in technique. Here we focus primarily on $R_{xx}$ traces; the
$R_{xy}$ data corroborate $R_{xx}$ and show corresponding plateaus.

\begin{figure}
\includegraphics[width=.4\textwidth]{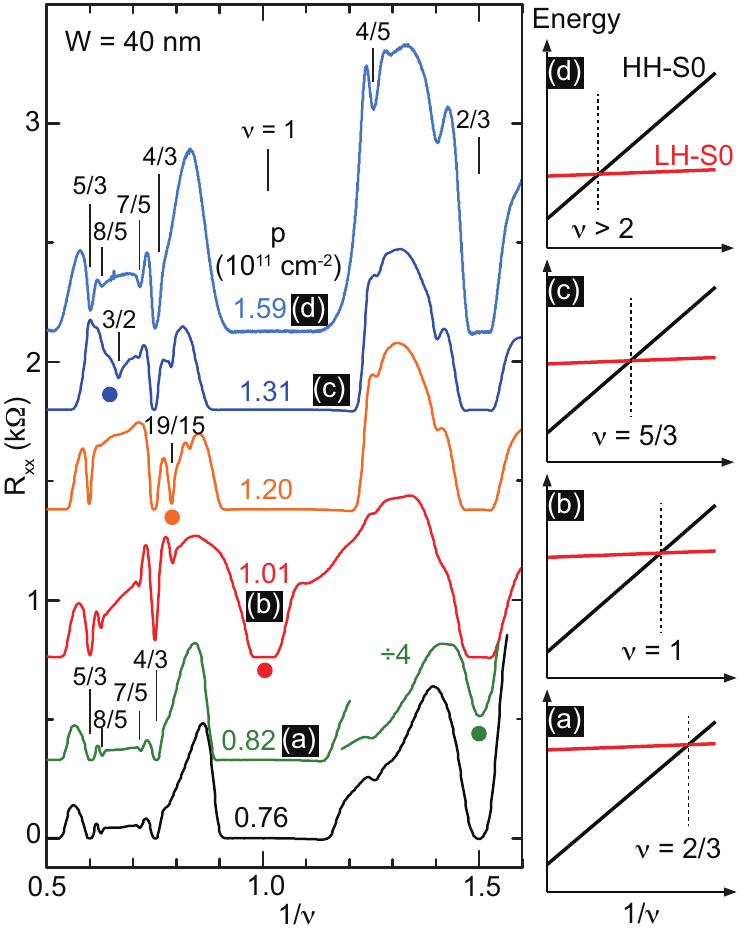}
\caption{$R_{xx}$ traces measured in a 2DHS confined to a 40-nm-QW at
  $\theta=0$ and different densities. Right panels show schematically
  the two lowest-energy LLs' energy vs $1/\nu$ at different densities,
  corresponding to the traces as marked.}
\end{figure}

\section{Experimental results at $\theta=0$}

We first describe data taken in a 2DHS confined to a 40-nm-wide QW as
a function of density at $\theta=0^{\circ}$. In Fig. 2, the QHS
transitions (marked by solid circles) which appear when two LLs are
nearly degenerate, can be seen moving from low to high $\nu$ as we
increase $p$. At $p=0.76$, we observe QHSs at the standard fillings,
similar to what is seen in systems where LLs from different subbands
are well-separated \cite{Jain.CF.2007}. The $\nu=2/3$ QHS becomes weak
at $p=0.82$ but is restored at higher $p$. The weakening of the
$\nu=1$ QHS at $p\simeq 1.01$ is evidenced by a profound narrowing of
its $R_{xx}$ plateau, and serves as direct evidence that the two
lowest-energy LLs are crossing at $\nu=1$ \cite{Note2, Maude.PRL.1996,
  Muraki.PRL.2001, Padmanabhan.PRL.2010, Liu.PRB.2011}.
At $p=1.20$, a strong $\nu=1$ QHS is restored, and a 2C QHS develops
at an unusual filling $\nu=19/15$ \cite{Manoharan.PRL.1997, Note3}.
The transition continues moving to higher $\nu$ at $p=1.31$. The
$\nu=5/3$ QHS disappears and another 2C QHS develops at $\nu=3/2$,
which is the particle-hole counterpart of the 2C $\nu=1/2$
($\Psi_{331}$) QHS \cite{Suen.PRL.1994}. In the top trace ($p=1.59$),
the 2DHS reverts back to 1C for $\nu<2$, exhibiting QHSs at standard
fillings. The above evolution of the QHSs, which implies a LL crossing
that moves from low $\nu$ to high $\nu$ as density is increased, is
consistent with previous observations and theoretical calculations
\cite{Liu.PRB.2014}.

The above LL crossing can be qualitatively understood in a simplified
picture (see the right panels in Fig. (2)). When confined to QWs,
because of their heavier mass in the $z$-direction, the heavy-hole
(HH) subband is lower in energy than the light-hole (LH) subband. But
the HHs have a smaller effective mass in the $xy$-plane than the LHs,
so the ground-state ($N=0$) LL of the HH symmetric subband, which we
refer to as HH-S0 for simplicity, increases faster in energy than the
LH-S0 LL as we sufficiently increase $B_{\perp}$, leading to a LL
crossing. In a more quantitative picture, the spin-orbit coupling
mixes the HH and LH subbands and LLs, and results in a more complex,
non-linear LL fan diagram. However, the crossing between the two
lowest-energy LLs is preserved in symmetric QWs
\cite{Liu.PRB.2014}. In our wide QW samples, the HH
  and LH subbands are close in energy, so the two levels cross at
  moderate $B_{\perp}$.

\begin{figure*}%
  \includegraphics[width=.8\textwidth]{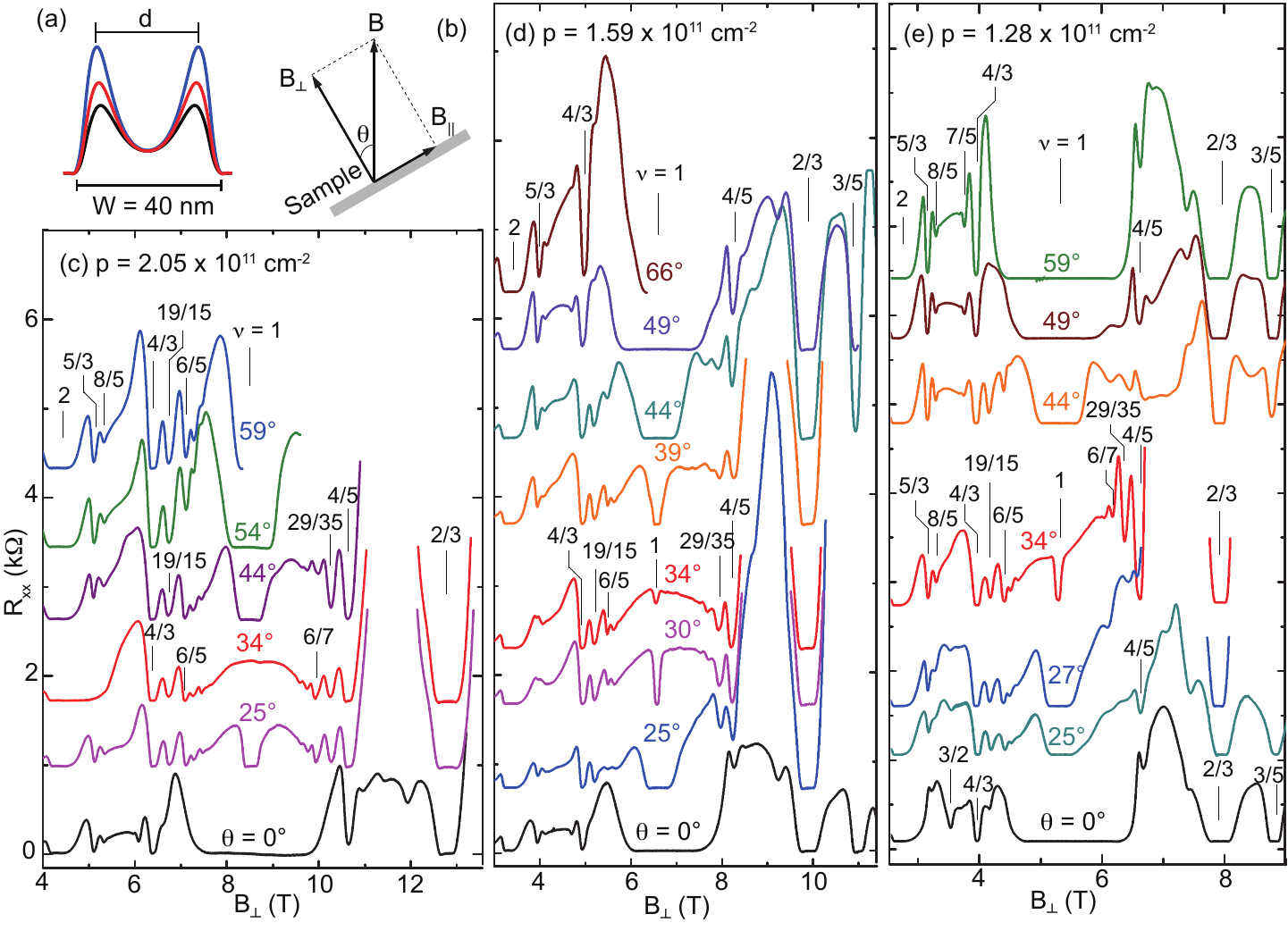}%
  \caption{(a) Self-consistently calculated charge distribution of the
    2DHS confined to the 40-nm-wide QW at densities $p=2.05$, 1.59 and
    1.28. (b) Experimental geometry. (c)-(e) $R_{xx}$ vs $B_{\perp}$
    traces measured at different $\theta$. In all panels, the QHS at
    $\nu=1$ is strong at $\theta=0$, disappears or weakens at
    $\theta\simeq 34^{\circ}$, and becomes strong again at larger
    $\theta$.}%
\end{figure*}

\section{Experimental results of finite $\theta$}

Data presented in Fig. 3 reveal that QHS transitions can also be
induced at a fixed density by varying $\theta$, but the behavior is
dramatically different. In Fig. 3(c), we show $R_{xx}$ vs $B_{\perp}$
traces measured at $p=2.05$ and different $\theta$. The density is
high so that the LH-S0 LL is well below the HH-S0 LL at
$\theta=0^{\circ}$ in the range $\nu<2$ (see Fig. 2(d)), and the 2DHS
exhibits 1C QHSs at standard fillings. At $\theta\simeq 34^{\circ}$,
the 2DHS becomes 2C in a large range of fillings $2/3<\nu<2$. This is
evinced by the development of insulating phases around $\nu=2/3$
(i.e., around $\nu=1/3$ for each component \cite{Santos.PRL.1992}),
the complete disappearance of the QHSs at $\nu=5/3$ and 1, as well as
the stabilization of QHSs at twice the standard fillings $\nu=4/3$,
6/5, 6/7, 2/3, and at unusual fillings such as $\nu=19/15$ and 29/35
\cite{Note3, Manoharan.PRL.1997}. At larger $\theta$, the $\nu=1$ and
5/3 QHSs reappear while many 2C QHSs remain, suggesting the two
lowest-energy LLs are separated by a small but finite energy
\cite{Note4}.

Figure 3(d) data taken at $p=1.59$ exhibit a more complete and
revealing evolution. The system is essentially 1C for
$\theta\lesssim 20^{\circ}$ and $\theta\gtrsim 44^{\circ}$, showing
strong QHSs at standard fillings \cite{Note4}.  It becomes 2C for
$\nu<2$ when $25^{\circ}\lesssim \theta \lesssim 44^{\circ}$,
exhibiting insulating phases flanking $\nu=2/3$ and 2C QHSs at
$\nu=19/5$, 6/5, 29/35, etc., while QHSs at $\nu=1$ and 5/3 become
weak and essentially disappear as $\theta$ approaches $34^{\circ}$.

Figure 3(e) shows traces taken at $p=1.28$ where, at $\theta=0$, the
LL crossing occurs near $\nu=3/2$, as evidenced by the stabilization
of the correlated, 2C QHS at $\nu=3/2$, and the absence of a QHS at
$\nu=5/3$. Similar to the data of Figs. 3(c) and (d), the system
becomes 2C near $\theta\simeq 34^{\circ}$ and 1C when
$\theta\gtrsim 49^{\circ}$. However, in contrast to Figs. 3(c) and (d)
data, the $\nu=1$ QHS becomes weak at $\theta=34^{\circ}$ but never
disappears. The fact that the system is 2C near $\nu=1$ suggests that
the $\nu=1$ QHS seen at $\theta\simeq 34^{\circ}$ in Fig. 3(d) is also
a 2C QHS; we will return to this later.

\section{Discussion}

The transition from 1C to 2C as a function of increasing $B_{||}$ has
been reported previously for \emph{electrons} confined to wide GaAs
QWs \cite{Manoharan.PRL.1997, Hasdemir.PRB.2015}. In such systems, the
coupling of $B_{||}$ to the orbital (out-of-plane) motion of electrons
renders the system progressively more bilayer-like at higher $B_{||}$
and quenches the energy separation between the $N=0$ LLs of the
symmetric and antisymmetric subbands, making them essentially
degenerate \cite{Manoharan.PRL.1997, Hasdemir.PRB.2015}. Further
increasing $B_{||}$ does not lift this degeneracy and the system
remains 2C at the highest $B_{||}$. This is very different from our
data shown in Figs. 3(d) and (e), where the 2DHS near $\nu=1$ becomes
2C only near $\theta\simeq 34^{\circ}$, but is 1C at smaller and
\textit{higher} $B_{||}$.

We attribute the evolution in Fig. 3 data to a $B_{||}$-induced LL
crossing \cite{Graninger.PRL.2011, Note5, Mueed.PRL.2015}. 
Unfortunately, no accurate calculations of LLs in the presence of both
$B_{\perp}$ and $B_{||}$ are available, particularly for 2DHSs with
multiband structure. The tilted-field geometry implies complicated
couplings between Landau harmonic oscillators from different subbands,
and makes numerical calculations extremely demanding. Qualitatively,
we can explain the crossing as follows. The densities of Fig. 3 data
are sufficiently large so that the LH-S0 level is lower than the HH-S0
level near $\nu=1$ at $B_{||}=0$ (Figs. 2(c) and (d)). Finite $B_{||}$
introduces additional confinement of the 2DHS in the $z$-direction,
raises the LH-S0 LL relative to the HH-S0 LL, and causes a crossing of
these levels at intermediate $\theta$ (see Fig. 4(d)).

The most remarkable feature of Fig. 3 data, however, is not the LL
crossing at an intermediate $\theta$. Rather, it is the behavior of
the 2DHS near the crossing angle, suggesting a very unusual
''pinning'' or near-pinning of the LLs in a very large range of $\nu$
(Fig. 1(b)). Note in Fig. 3 that at a given density the system
exhibits 2C behavior in the entire range of $\nu<4/3$ at
$\theta\simeq 34^{\circ}$. This is very different from the $\theta=0$
data of Fig. 2 where the LL crossing features for any given density
appear near a specific $\nu$ which moves from low to high values as
the density is increased. Moreover, in Fig. 3 the angle
$\theta\simeq 34^{\circ}$ at which the 2DHS becomes 2C appears to be
independent of the 2DHS density. In other 2DHS samples, confined to
QWs with $W$ ranging from 35 to 50 nm, we have observed similar
phenomena as in Fig. 3 at the same $\theta\simeq 34^{\circ}$. This
independence of the 2C behavior on $\nu$, $p$, and $W$ at this
critical angle is astonishing, and demands a theoretical explanation.

\begin{figure}[htb]%
\includegraphics[width=.45\textwidth]{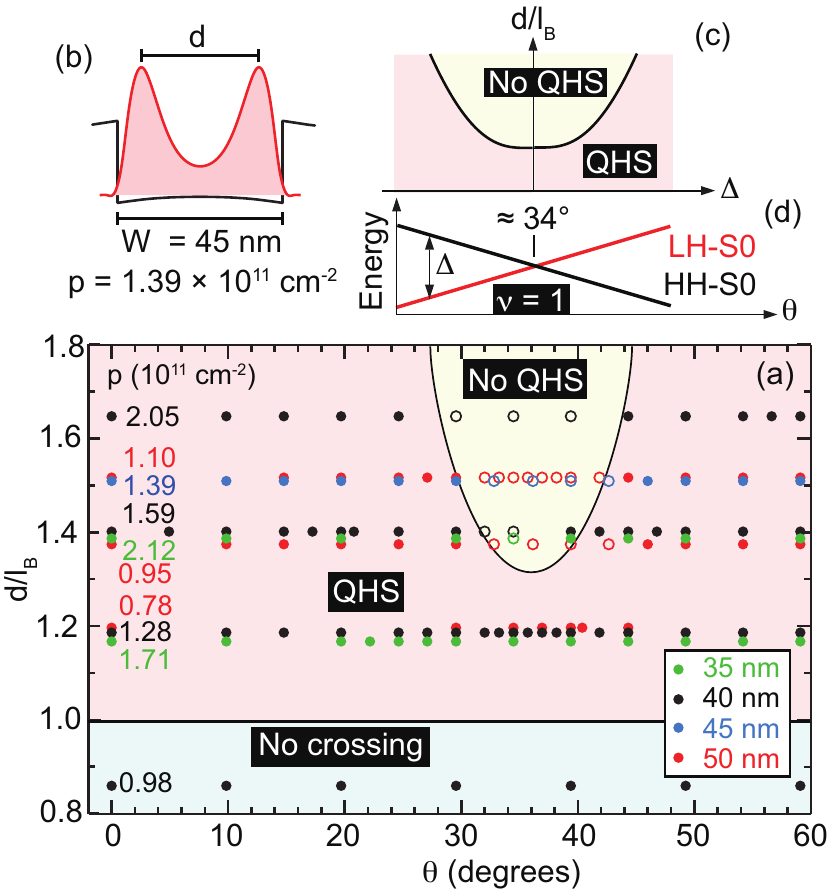}%
\caption{(a) Phase diagram for the stability of QHS at $\nu=1$ as a
  function of the tilting angle $\theta$ and $d/l_B$. The solid (open)
  symbols mark the presence (absence) of a QHS at $\nu=1$. In narrow
  QWs and at low density, no crossing is seen as a function of
  $\theta$, shown as the blue region. Once $d/l_B\gtrsim 1.0$, a
  crossing occurs near $\theta\simeq 34^{\circ}$. At the crossing, a
  QHS appears at $\nu=1$ if $d/l_B\lesssim 1.3$, consistent with the
  $\Psi_{111}$ state. (b) Calculated charge distribution for a
  45-nm-wide QW with $p=1.39$ showing the interlayer distance
  $d$. (c)-(d) Schematic phase and LL diagrams at $\nu=1$ showing how
  the LL separation $\Delta$ increases as $\theta$
  deviates from $\simeq 34^{\circ}$.}
\end{figure}

The evolution of the QHS at $\nu=1$ is also very intriguing. As seen in
Fig. 3, it disappears completely at $\theta\simeq 34^{\circ}$ when
$p=2.05$ but only becomes weak at $p=1.28$. In Fig. 4(a) we summarize
our results for many 2DHSs, illustrating the conditions for the
stability of the $\nu=1$ QHS. Data are shown as a function of $\theta$
and $d/l_B$, which compares the interlayer ($e^2/4\pi\epsilon d$) and
intra-layer ($e^2/4\pi\epsilon l_B$) correlations and is widely used
to characterize bilayer QHSs \cite{Murphy.PRL.1994, Spielman.PRL.2000,
  Kellogg.PRL.2004, Tutuc.PRL.2004, Tutuc.PRB.2005, Lay.PRB.1994,
  Eisenstein.ARCMP.2014, Note6}. 
Figure 4(a) shows that no LL crossing at $\nu=1$ can be induced via
tilting if $d/l_B\lesssim 1.0$, and the $\nu=1$ QHS is always
strong. When $d/l_B \gtrsim 1.0$, at $\nu=1$, the LH-S0 level is lower
than the HH-S0 level at $\theta=0$, and the two levels cross at
$\theta\simeq 34^{\circ}$; see Fig. 4(d). At the crossing, we observe
a QHS at $\nu=1$ if $d/l_B\lesssim 1.3$, and the ground state becomes
compressible if $d/l_B\gtrsim 1.3$. 

The $d/l_B\lesssim 1.3$ condition for the stability of the $\nu=1$ QHS
at the crossing, and the fact that the 2DHS is 2C at nearby fillings,
suggest that it is a 2C QHS with strong interlayer correlations,
likely the $\Psi_{111}$ state reported in GaAs bilayer electron
\cite{Murphy.PRL.1994, Spielman.PRL.2000, Kellogg.PRL.2004,
  Eisenstein.ARCMP.2014} or hole \cite{Tutuc.PRL.2004, Tutuc.PRB.2005}
systems confined to double QWs. In those systems, when the lowest LLs
from different subbands are degenerate, the $\nu=1$ QHS is stable at
$d/l_B\lesssim 2$, and turns into a compressible state if $d/l_B$
becomes large \cite{Murphy.PRL.1994, Eisenstein.ARCMP.2014}. Also note
that in our experiments the energy separation between the two crossing
LLs increases as $\theta$ deviates from $\simeq 34^{\circ}$ (see
Fig. 4(d)). We show in Fig. 4(c) a schematic ''phase diagram'' for the
stability of the $\nu=1$ QHS as functions of $\Delta$ and $d/l_B$. The
resemblance of Fig. 4(c) to the phase diagram of $\nu=1$ QHS in
double QWs \cite{Murphy.PRL.1994, Eisenstein.ARCMP.2014} is
striking. We emphasize that in our experiments, we are essentially
tuning $\Delta$ through zero as we tilt the sample near
$\theta\simeq 34^{\circ}$; see Figs. 4(d).

In conclusion, 2DHSs confined to wide GaAs QWs and with sufficiently
high density, reveal an unusual crossing of the two lowest-energy LLs
near $\nu=1$ as we tilt the sample in magnetic field. It appears at a
magic angle $\theta\simeq 34^{\circ}$, essentially independent of the
QW width, density, or $B_{\perp}$ (filling), suggesting a pinning of
the LLs near the crossing. The crossing and the pinning likely stem
from the complex interplay of the heavy- and light-hole LLs in
$B_{||}$, and should stimulate further theoretical investigation. Near
this angle, the 2DHS becomes 2C at $\nu<2$ and, if $d/l_B$ is small,
exhibits a $\nu=1$ QHS, consistent with a correlated, 2C, $\Psi_{111}$
state.

\begin{acknowledgments}
  We acknowledge the DOE BES (DE-FG02-00-ER45841) grant for
  measurements, and the NSF (Grants DMR-1305691 and MRSEC
  DMR-1420541), the Gordon and Betty Moore Foundation (Grant
  GBMF4420), and Keck Foundation for sample fabrication and
  characterization. A portion of this work was performed at the NHMFL,
  which is supported by the NSF Cooperative Agreement No. DMR-1157490,
  the State of Florida, and the DOE. We thank S. Hannahs, G. E. Jones,
  T. P. Murphy, E. Palm, A. Suslov, and J. H. Park for technical
  assistance. We are indebted to R. Winkler for illuminating
  discussions and providing the self-consistently calculated charge
  distribution.
\end{acknowledgments}

\bibliographystyle{apsrev4-1}
\bibliography{../bib_full}

\end{document}